\documentclass[a4paper,twocolumn,citeautoscript,prl]{revtex4-1}

\usepackage{amsmath,bm}
\usepackage[pdftex]{graphicx,color}
\usepackage[utf8]{inputenc}

\begin{document}

\title{Octave-spanning tunable parametric oscillation in crystalline Kerr microresonators}

\author{Noel Lito B. Sayson$^{1,2,3}$}
\author{Toby Bi$^{1,2}$}
\author{Vincent Ng$^{1,2}$}
\author{Hoan Pham$^{1,2}$}
\author{Luke S. Trainor$^{1,4}$}
\author{Harald G. L. Schwefel$^{1,4}$}
\author{St\'ephane Coen$^{1,2}$}
\author{Miro Erkintalo$^{1,2}$}
\author{Stuart G. Murdoch$^{1,2}$}
\email{s.murdoch@auckland.ac.nz}

\affiliation{$^1$The Dodd-Walls Centre for Photonic and Quantum Technologies, New Zealand}
\affiliation{$^2$Department of Physics, University of Auckland, Auckland 1010, New Zealand}
\affiliation{$^3$Physics Department, Mindanao State University - Iligan Institute of Technology, Tibanga, 9200 Iligan City, Philippines}
\affiliation{$^4$Department of Physics, University of Otago, Dunedin 9016, New Zealand}

\begin{abstract}
\noindent Parametric nonlinear optical processes allow for the generation of new wavelengths of coherent electromagnetic radiation. Their ability to create radiation that is widely tunable in wavelength is particularly appealing, with applications ranging from spectroscopy to quantum information processing. Unfortunately, existing tunable parametric sources are marred by deficiencies that obstruct their widespread adoption. Here we show that ultrahigh-Q crystalline microresonators made of magnesium fluoride can overcome these limitations, enabling compact and power-efficient devices capable of generating clean and widely-tunable sidebands. We consider several different resonators with carefully engineered dispersion profiles, achieving hundreds of nanometers of sideband tunability in each device when driven with a standard low-power laser at 1550 nm. In addition to direct observations of discrete tunability over an entire optical octave from 1083 nm to 2670 nm, we record signatures of mid-infrared sidebands at almost 4000 nm. The simplicity of the devices considered -- compounded by their remarkable tunability -- paves the way for low-cost, widely-tunable sources of electromagnetic radiation.
  \end{abstract}
\maketitle

\section{Introduction}
\noindent High-Q optical microresonators have emerged over the last two decades as a revolutionary new platform for nonlinear optics~\cite{vahala_optical_2003, strekalov_nonlinear_2016}. Their ultra-high finesse and small modal volume enables nonlinear interactions to be driven with unprecedented efficiency, permitting e.g. harmonic generation at microwatt input levels~\cite{furst_naturally_2010}. Resonators dominated by third-order Kerr nonlinearities have attracted particular attention~\cite{kippenberg_kerr-nonlinearity_2004, savchenkov_low_2004, moss_new_2013}, and notably allowed for the development of chip-scale generators of coherent optical frequency combs~\cite{delhaye_optical_2007, kippenberg_microresonator-based_2011, webb_experimental_2016, cole_soliton_2017, pasquazi_micro-combs:_2017, xue_second-harmonic-assisted_2017, spencer_optical-frequency_2018, kippenberg_dissipative_2018, stern_battery-operated_2018}. The application of such ``Kerr'' frequency combs now constitutes a major subject of research, with impressive demonstrations reported in contexts such as spectroscopy~\cite{suh_microresonator_2016, yu_silicon-chip-based_2018, dutt_-chip_2018}, optical ranging~\cite{suh_soliton_2018, trocha_ultrafast_2018}, and telecommunications~\cite{marin-palomo_microresonator-based_2017, fulop_high-order_2018}.

The nonlinear optical phenomenon that underlies the formation of Kerr combs is known as four-wave-mixing (FWM). In addition to comb formation, FWM can also be harnessed for the generation of isolated pairs of new optical frequencies that exhibit large frequency-shifts from the pump and that can be widely tuned by small changes in the pump wavelength~\cite{lin_phase_1981}. While such Kerr parametric sources have been extensively studied using optical fibres as the nonlinear medium~\cite{harvey_scalar_2003, pitois_experimental_2003, wong_high-conversion-efficiency_2007, bessin_modulation_2017}, the low nonlinearity and finesse of silica fibre-based systems imposes severe limitations on power efficiency and spectral coverage. High-Q Kerr microresonators are ideally positioned to overcome these limitations, offering an intriguing possibility for the development of a new type of widely-tunable optical source.

Several pioneering studies have demonstrated the generation of large frequency shift parametric sidebands in Kerr nonlinear optical microresonators~\cite{liang_miniature_2015, matsko_clustered_2016, sayson_widely_2017, fujii_third-harmonic_2017, huang_quasi-phase-matched_2017}. The ability to tune the wavelengths of the generated sidebands has, however, remained elusive, with the majority of studies reporting little~\cite{huang_quasi-phase-matched_2017} or no~\cite{liang_miniature_2015, matsko_clustered_2016, fujii_third-harmonic_2017} tunability. The largest tuning range reported to date was observed in a fused silica microsphere, where sidebands tunable from 1200~nm to 1900~nm were generated~\cite{sayson_widely_2017}. However, the high attenuation of fused silica at wavelengths beyond 1900~nm fundamentally prevents further wavelength-scaling of this type of oscillator, obstructing access to the spectroscopically rich mid-infrared. Moreover, the broad Raman gain of fused silica invariably results in the concomitant generation of numerous parasitic spectral components, which together with the unwieldy microsphere geometry critically diminishes the attractiveness of this system. To fully unleash the potential of microresonator Kerr parametric oscillators, a platform capable of overcoming these deficiencies is needed.

\begin{figure*}[!t]	
\centering
		\includegraphics[width=\linewidth]{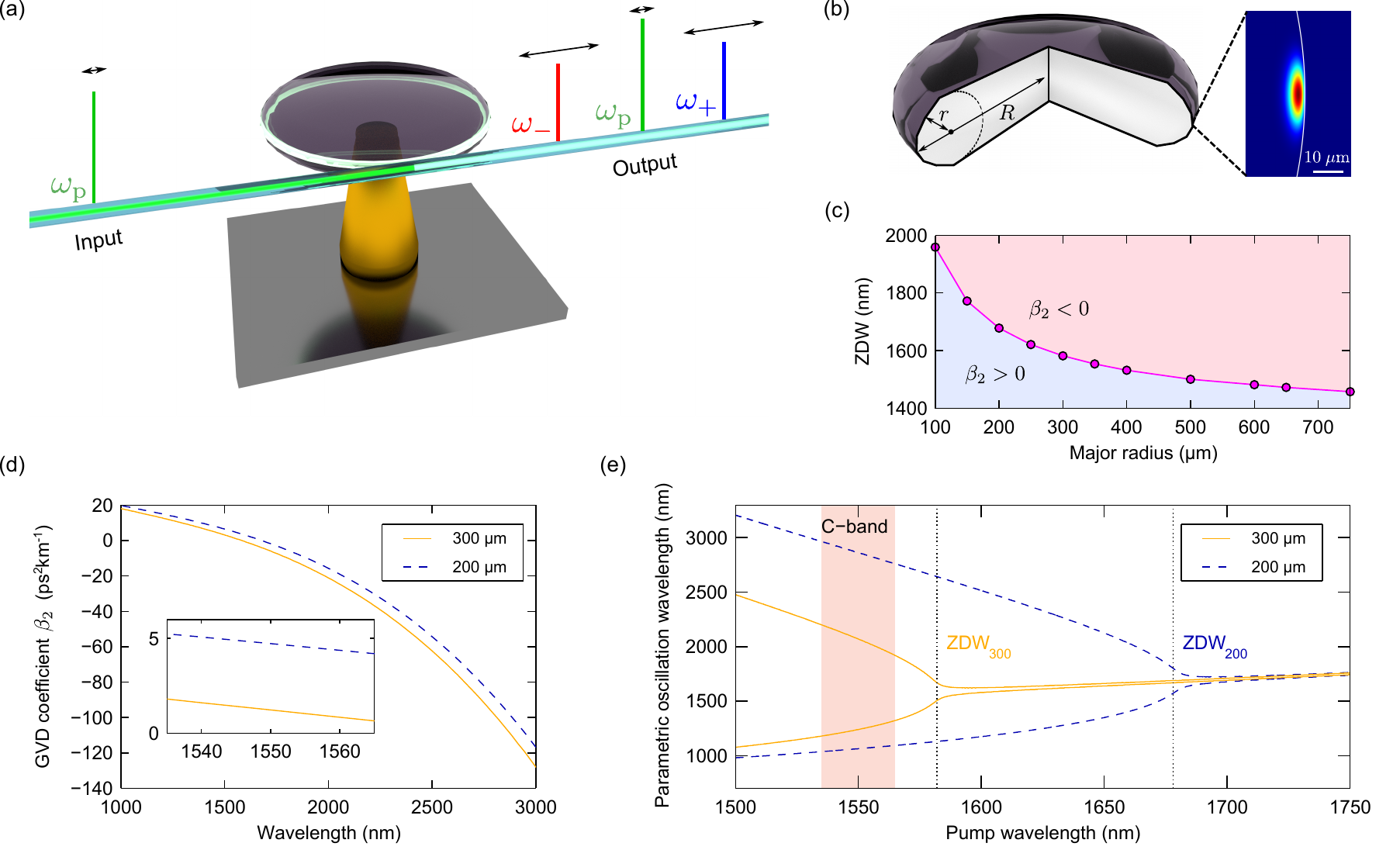}
		\caption{\textbf{Scheme description and resonator modelling.} (a) Schematic illustration of widely-tunable parametric oscillation in a Kerr microresonator. A cw beam with an angular frequency $\omega_\mathrm{p}$ coupled into a high-Q Kerr microresonator results in the generation of symmetrically detuned sidebands with frequencies $\omega_\pm$ via degenerate FWM. If the resonator exhibits appropriate dispersion characteristics, small changes in the pump frequency map into wide sideband tunability. (b) Disk geometry assumed in the modelling of resonator dispersion. Inset shows the field distribution of the fundamental TE mode of a resonator with a major radius $R = 300~\mathrm{\mu m}$. (c) Modelled zero-dispersion wavelength (ZDW) as a function of the resonator major radius $R$ for a constant minor radius. Wavelengths above (below) the ZDW exhibit anomalous (normal) GVD. (d) Modelled wavelength-dependence of the GVD coefficient $\beta_2$ for two resonators with $R = 300~\mathrm{\mu m}$ and $R = 200~\mathrm{\mu m}$. Inset shows a zoom around C-band wavelengths. (e) Theoretically predicted parametric oscillation wavelengths as a function of the pump wavelength for the two resonators considered in (d). Dotted vertical lines indicate the ZDWs of the resonators. The phase-matching curve in (e) uses values typical to our experiment: $\gamma = 1~\mathrm{W^{-1}km^{-1}}$, $P = 100~\mathrm{W}$, and $\delta_0 = 0~\mathrm{rad}$. All calculations assume a resonator minor radius of $r = 130~\mathrm{\mu m}$.}
		\label{fig1}
\end{figure*}

Here, we present experimental demonstrations of widely-tunable parametric oscillation in crystalline Kerr microdisk resonators made of magnesium fluoride (MgF$_2$). Because MgF$_2$ remains transparent to wavelengths up to 6000~nm, the material loss restriction encountered in~\cite{sayson_widely_2017} is lifted. In addition, thanks to the narrowband Raman gain spectrum characteristic of crystalline materials, pure Kerr FWM signals can be generated. We experimentally consider several different resonators with dispersion profiles carefully engineered to give access to different wavelength regions, observing clean and low-noise sidebands that can be tuned by hundreds of nanometers with a standard C-band telecommunications laser as the pump. Combining three distinct resonators, we realize over an octave of narrowband, continuous-wave (cw) tunable output ranging from 1083~nm to 2670~nm -- at an input pump power of only 100~mW. Moreover, we observe signatures of spectral components at wavelengths as high as 3800~nm, thus demonstrating the scheme's potential for mid-infrared (mid-IR) applications. Our work paves the way for future low-cost, low-power, widely-tunable sources of electromagnetic radiation based on Kerr microresonators.

\section{Results}
\subsection{Overall scheme and phase-matching}

\noindent Figure~\ref{fig1}(a) shows a schematic illustration of our experiment. A cw laser with frequency $\omega_\mathrm{p}$ is coupled into a high-Q MgF$_2$ microdisk resonator, resulting in the generation of large-frequency shift signal and idler sidebands through degenerate FWM. Energy conservation dictates that the sidebands are symmetrically detuned around the pump, $\omega_\pm = \omega_\mathrm{p}\pm\Omega$. To enable wide tunability, we operate in a regime where small adjustments of the pump frequency give rise to very large changes in the sideband frequency shift $\Omega$. This is achieved by phase-matching the nonlinear wave mixing process through higher-order dispersion~\cite{harvey_scalar_2003, sayson_widely_2017}. Assuming all interacting waves share the same mode family, the phase-matching condition can be approximately written as [see Methods]
\begin{equation}
\label{PM}
\frac{\beta_2\Omega^2 L}{2} + \frac{\beta_4\Omega^4 L}{24} + 2\gamma P L - \delta_0 = 0,
\end{equation}
where $\beta_2$ and $\beta_4$ are respectively the second- and fourth-order group-velocity dispersion (GVD) coefficients of the resonator mode family under study (evaluated at the pump frequency), $L$ is the resonator circumference, $P$ is the intracavity power at the pump frequency, $\gamma$ is the nonlinear interaction coefficient of the mode family, and $\delta_0$ is the phase detuning between the pump field and the closest linear cavity resonance.

Inspection of Eq.~\eqref{PM} reveals that large-frequency shift, widely tunable sidebands can be expected when the pump experiences normal group-velocity dispersion ($\beta_2>0$) and a negative fourth-order dispersion coefficient ($\beta_4<0$). In this regime, the negative fourth-order dispersion can compensate for the positive second-order dispersion at large frequency shifts $\Omega\approx\sqrt{-12\beta_2/\beta_4}$; for typical parameters, shifts of the order of 10 -- 100 THz can be expected \cite{lin_phase_1981, harvey_scalar_2003, pitois_experimental_2003, wong_high-conversion-efficiency_2007, sayson_widely_2017}. Moreover, the precise frequency shift can be tuned over wide regions by adjusting the values of the dispersion coefficients $\beta_2$ and $\beta_4$, which can be readily achieved through small changes of the pump frequency [see Methods].

\subsection{Resonator dispersion}

\noindent The resonators used in our experiments are made of MgF$_2$, which exhibits a very low loss over the target spectral region ($1000 - 4000$~nm) and is amenable to fabrication using established single-point turning techniques~\cite{grudinin_properties_2016,sedlmeir_experimental_2013, wang_mid-infrared_2013}. We wish to drive the resonators using a pump source in the telecommunications C-band ($1530 - 1565$~nm) so as to take advantage of the low-cost, high-quality lasers and components available at these wavelengths. Unfortunately, the zero-dispersion ($\beta_2=0$) wavelength (ZDW) of bulk MgF$_2$ is located at about 1300~nm, with normal dispersion at shorter wavelengths~\cite{lin_dispersion_2015}. To achieve normal dispersion at 1550~nm, as required for large-frequency shift sideband generation, additional waveguide dispersion is necessary to shift the ZDW to longer wavelengths.

We first performed detailed modelling using a commercial finite-element software (COMSOL Multiphysics) to identify resonator dimensions that satisfy the phase-matching conditions required for widely-tunable parametric oscillation [see also Methods]. We consider a microdisk resonator geometry characterised by major and minor radii $R$ and $r$, respectively, as shown in Fig.~\ref{fig1}(b). Our modelling reveals that, for the range of parameters accessible with our fabrication ($R \approx 100 - 500~\mathrm{\mu m}$ and $r \approx 50 - 250~\mathrm{\mu m}$), the ZDW can be straightforwardly increased by reducing the resonator's major radius, whereas changing the minor radius has a much smaller effect. Illustrative results are shown in Fig.~\ref{fig1}(c), where we plot the ZDW of the fundamental TE mode as a function of the major radius $R$  for a minor radius $r = 130~\mathrm{\mu m}$ (all the modelling results shown use this value). The results show that the ZDW can be shifted to 1550~nm by reducing the major radius $R\lesssim500~\mathrm{\mu m}$, thus alluding to the feasibility of widely tunable parametric oscillation with a C-band pump.

Figure~\ref{fig1}(d) shows the modelled wavelength-dependence of the $\beta_2$ coefficient for two resonators with major radii of 300~$\mu$m and 200~$\mu$m. Two important points are immediately discernable. First, both resonators exhibit normal dispersion over our intended range of pump wavelengths around 1550~nm. Second, the curvatures of $\beta_2$ are negative, implying $\beta_4<0$. These resonators thus fulfil both requirements for achieving widely-tunable parametric oscillation. Figure~\ref{fig1}(e) shows the phase-matched sideband wavelengths predicted by Eq.~\eqref{PM} as a function of the pump wavelength for these two resonators [other parameters listed in the caption]. As can be seen, pumping in the normal dispersion regime (below the ZDW) is expected to result in parametric sidebands with very large frequency shifts that change dramatically in response to small changes in the pump wavelength. For the 300~$\mu$m resonator (solid orange curve), the tuning range predicted as the pump is tuned by 80~nm from 1500~nm to 1580~nm is well in excess of an octave ($1000 - 2500$~nm). Since our experimental implementation is, however, limited to pump wavelengths in the C-band, octave-tunability is not accessible with a single resonator. We lift this limitation by fabricating resonators with different major radii, which provides a simple route to control the position of the ZDW and hence the entire phase-matching curve. Indeed, as shown by the dashed blue curve in Fig.~\ref{fig1}(e), a resonator with a major radius of 200~$\mu$m is predicted to give access to significantly larger frequency shifts for the same range of pump wavelengths compared to the 300~$\mu$m resonator.

\begin{figure*}[!t]	
\centering
		\includegraphics[width=\linewidth]{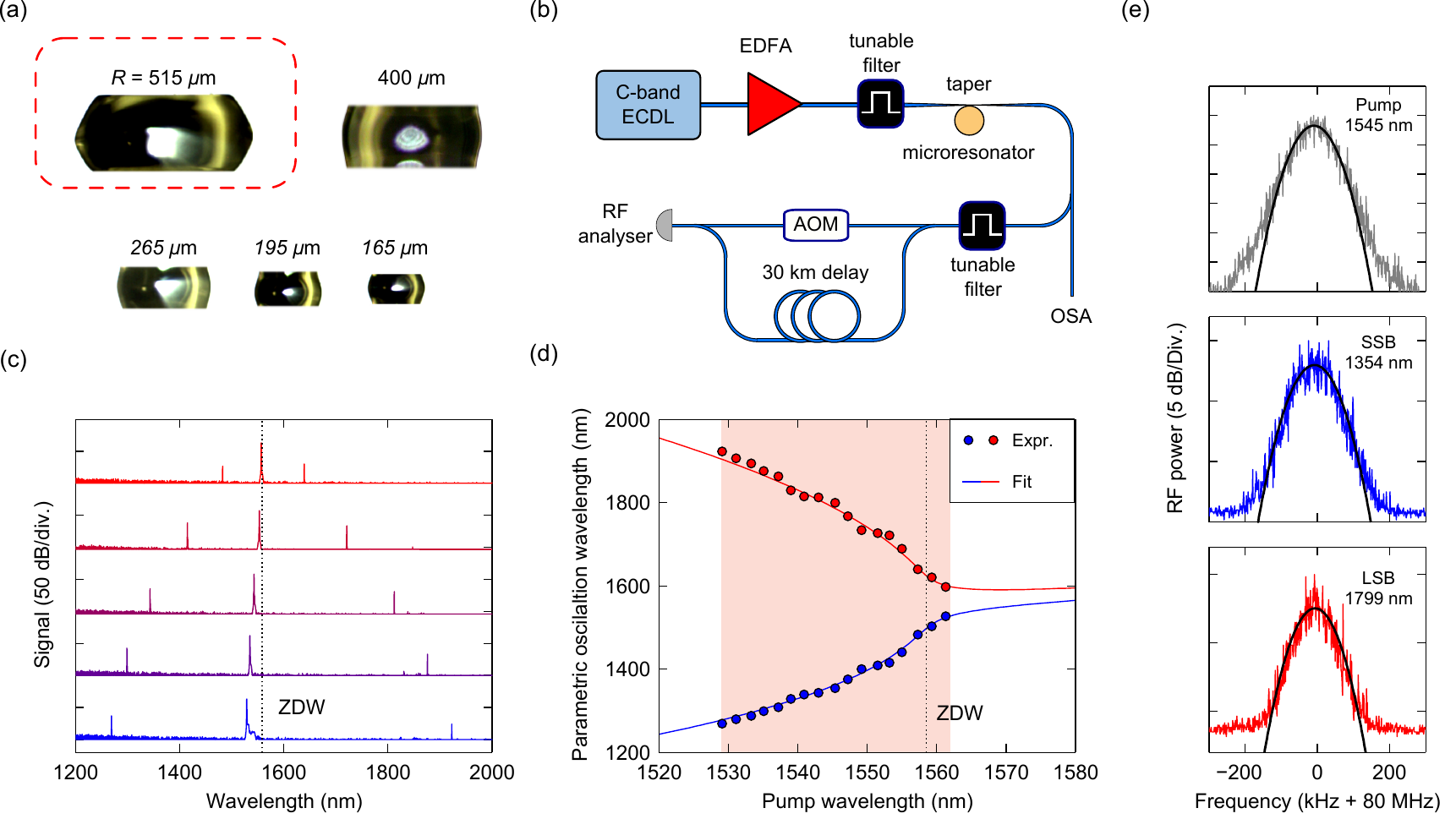}
		\caption{\textbf{Experimental setup and illustrative results.} (a) Photographs of the resonators used in our experiments with major radii $R$ as indicated. (b) Schematic illustration of the experimental setup. OSA, optical spectrum analyser; AOM, acousto-optic modulator. (c--e) Experimental results obtained for a resonator with a major radius of $R = 515~\mathrm{\mu m}$ [highlighted with a dashed rectangle in (a)] and for a pump power of 100~mW. (c) Measured spectra at the resonator output for five different pump wavelengths $\lambda_\mathrm{p}$ (from top to bottom, $\lambda_\mathrm{p}$ = 1557, 1553, 1543, 1535, 1529~nm). (d) Measured sideband wavelengths as a function of pump wavelength (solid circles), together with a theoretical fit to the phase-matching Eq.~\eqref{PM} (solid curves). (e) Results from self-heterodyne measurements for the pump line, the short-wavelength sideband (SSB), and the long-wavelength sideband (LSB), with the pump wavelength set to $\lambda_\mathrm{p} = 1545~\mathrm{nm}$. The linewidths of the pump and the sidebands are estimated to be about 75~kHz and 106~kHz, respectively, based on Gaussian fits to the measured data (solid black curves).}
		\label{fig2}
\end{figure*}

\subsection{Experimental setup and results}

\noindent Following the analysis above, we fabricated five MgF$_2$ microdisk resonators with progressively decreasing dimensions [see Fig.~\ref{fig2}(a)] so as to access a wide range of sideband wavelengths. The resonators are first shaped on an air bearing lathe using diamond point turning, and then mechanically polished with diamond abrasives to achieve high finesse. After polishing, the major radii of the five resonators were measured to be 515, 400, 265, 190, and 165~$\mu$m, with the minor radii about one third of the corresponding major radius. The measured finesse of each resonator is about $50{,}000$, corresponding to ultra-high Q-factors ranging from $0.5~\times 10^{8}$ to $1.5~\times 10^{8}$.

We drive the resonators with a standard C-band external cavity diode laser (ECDL) that is amplified with an Erbium-doped optical amplifier (EDFA) [see Fig.~\ref{fig2}(b) and Methods]. The ensuing pump light has an average power of about 100~mW and a wavelength that is continuously tunable between 1530~nm and 1570~nm. The pump light is coupled to the microresonators using a silica optical fibre taper with a waist diameter of about one micron. 

Focusing first on results obtained for the 515~$\mu$m resonator, Fig.~\ref{fig2}(c) shows experimentally measured spectra at the taper output for five different pump wavelengths (corresponding to different driven cavity modes from the same mode family). At each wavelength, the initially blue-detuned pump laser is tuned into resonance until parametric oscillation is observed (thermal locking allows for stable operation~\cite{carmon_dynamical_2004}). As the pump wavelength is decreased, the observed parametric frequency shifts rapidly increase, as expected for phase-matching via higher-order dispersion. Each of the measured spectra comprise solely of a single pair of FWM sidebands, with no evidence of the strong stimulated Raman components observed in earlier experiments with silica microspheres~\cite{sayson_widely_2017}. We attribute this to the narrow Raman gain spectrum of crystalline MgF$_2$ which is typically not resonant with the mode family under investigation (unless specifically engineered to be so).

We repeat the spectral measurements for a range of pump wavelengths in the C-band, advancing the pump wavelength by about three free-spectral ranges for each data point. Figure~\ref{fig2}(d) shows the wavelengths of the parametric sidebands extracted from each measurement. As can be seen, just 30~nm of pump tuning in the C-band yields more than 650~nm of signal tunability, with sidebands ranging from 1270~nm to 1920~nm observed. Also shown as the solid lines in Fig.~\ref{fig2}(d) is a theoretical fit to the phase-matching condition given by Eq.~\eqref{PM}, with the dispersion coefficients $\beta_2$, $\beta_3$, and $\beta_4$ treated as fitting parameters [see Methods]. Agreement is outstanding, confirming that the FWM process is phase-matched via higher-order dispersion. From this fit, we are also able to extract the dispersion coefficients and the ZDW of the mode family under study [the extracted ZDW of 1558~nm is highlighted with a dashed line in Fig.~\ref{fig2}(c) and (d)]. We have also performed delayed self-heterodyne measurements [Methods] to confirm the narrow linewidths of selected sidebands [Fig.~\ref{fig2}(e)]. We observe the sideband linewidths to be about 100~kHz, which is comparable to the original pump linewidth of about 70 kHz. These measurements show that the parametric sidebands share the noise characteristics of the pump, which is an important feature for applications requiring signals with narrow optical linewidth.

\begin{figure*}[!t]	
\centering
		\includegraphics[width=\linewidth]{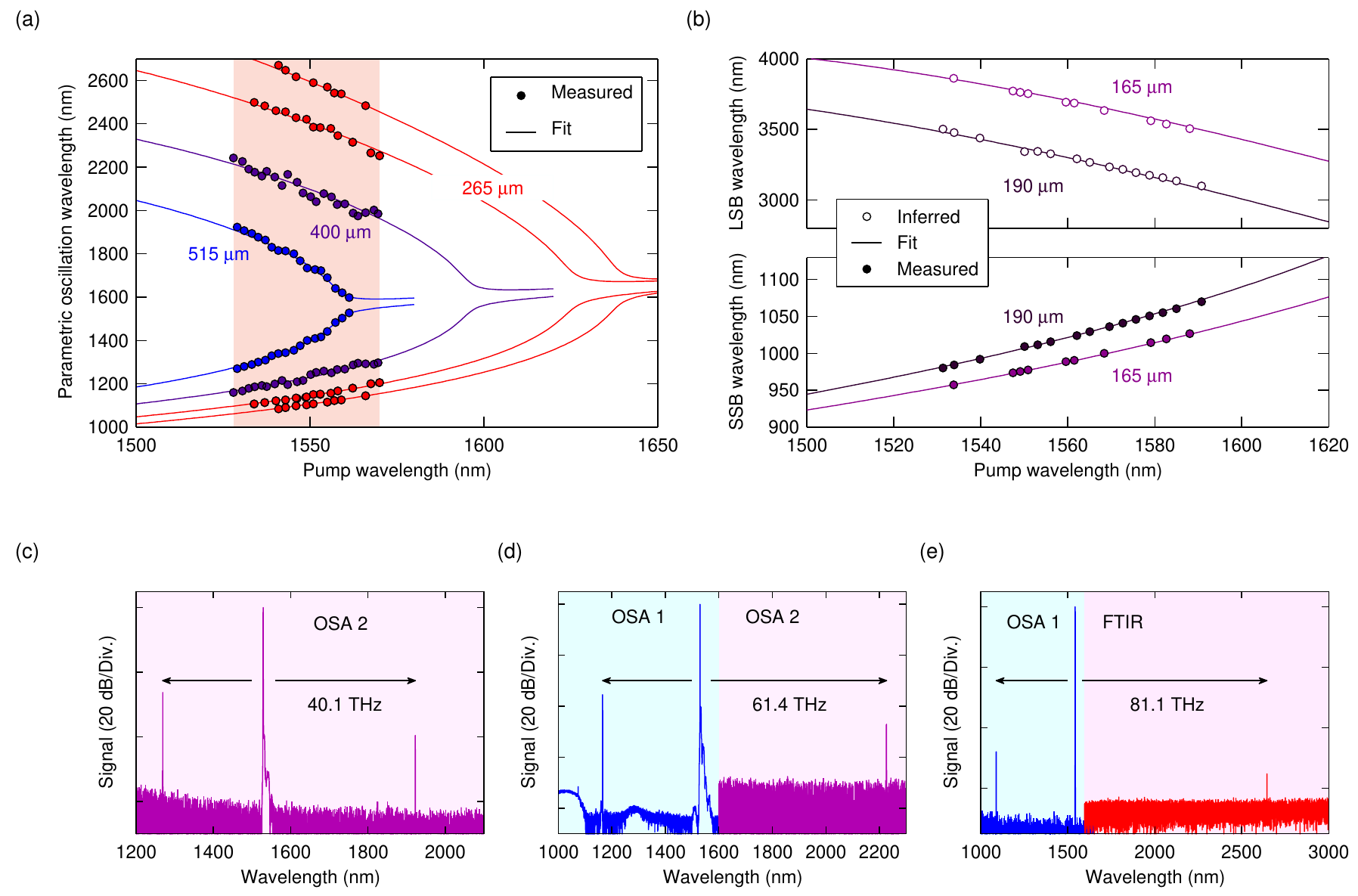}
		\caption{\textbf{Experimental observations of octave-tunability and signatures of mid-infrared sidebands.} (a) Measured sideband wavelengths as a function of the pump wavelength in three resonators with major radii of $R = 515~\mathrm{\mu m}$, $R = 400~\mathrm{\mu m}$, and $R = 265~\mathrm{\mu m}$.  The shaded area highlights the entire range of pump tunability (1530~nm - 1570~nm) while the solid curves show theoretical fits to the phase-matching curve defined by Eq.~\eqref{PM}. Note that, for the $265~\mathrm{\mu m}$ resonator, results from separate experiments corresponding to two different mode families are shown. (b) Solid circles show measured short-wavelength sidebands (SSB) while the open circles show corresponding inferred long-wavelength sidebands (LSB) in two resonators with major radii of $R = 190~\mathrm{\mu m}$ and $R = 165~\mathrm{\mu m}$. Solid curves show theoretical fits. (c--e) Typical spectra measured from each of the three resonators considered in (a): (c) R = 515~$\mu$m, (d) R = 400~$\mu$m, and (e) R = 265~$\mu$m. Two different OSAs were used to measure the combined spectrum shown in (d), while the combined spectrum in (e) was obtained with one OSA and an FTIR spectrometer. In (c) and (d), the pump wavelength was set to 1530~nm, while in (e) the pump wavelength was 1543~nm. All measurements were obtained with a pump power of 100~mW.
}
		\label{fig3}
\end{figure*}

The results reported in Fig.~\ref{fig2} show that a MgF$_2$ microdisk resonator with a 515~$\mu$m major radius can deliver parametric sidebands with over 650~nm of tunability when driven with a standard C-band pump laser. Remarkably, we find that significantly larger frequency shifts can be achieved by using resonators with smaller major radii (hence, larger ZDWs). Following a similar procedure as for the 515~$\mu$m resonator, we measure the wavelengths of the parametric sidebands generated in each of the five resonators fabricated. Here the extreme frequency shifts achieved in our resonators necessitates the use of two different optical spectrum analyzers (the first measures from 600 nm to 1750 nm, whilst the second measures from 1200 nm to 2400 nm) as well as a Fourier-transform infrared spectrometer (FTIR, used to measure signals above 2400~nm). Figure~\ref{fig3}(a) shows the experimental results obtained for three of our largest devices (with major radii of 515, 400, and 265~$\mu$m), together with fits to the theoretical phase-matching curves defined by Eq.~\eqref{PM}. Note that, for the 265~$\mu$m resonator, we identified two distinct mode families producing large frequency-shift sidebands: Fig.~\ref{fig3}(a) shows the tuning curves pertaining to both families as obtained from independent experiments where the pump laser is separately tuned to the spectrally distinct resonances of the respective mode families. We also note that detection of sidebands with wavelengths larger than about 2300~nm required us to reduce the length of the silica fibre after the coupling region to only 5~cm in an effort to minimise the very large attenuation of silica glass at these wavelengths.

\looseness=-1The frequency shifts reported in Fig.~\ref{fig3}(a) are remarkable. Hundreds of nanometres of sideband tunability is achieved in each resonator, with the combined output of the three devices representing more than an octave of quasi-continuous tunability from 1083~nm to 2670~nm. Yet, our experiments signal that even larger frequency shifts can be realised by further reducing the resonator dimensions. Indeed, as shown in Fig.~\ref{fig3}(b), the shortest wavelengths measured in our smallest resonators (with major radii of 190 and 165~$\mu$m) are well below 1000~nm (down to 957~nm), implying the generation of mid-IR wavelengths well above 3000~nm (up to 3860~nm). (Note that an additional L-band laser source was used here to interrogate the tuning curves over an extended range of pump wavelengths [see Methods].) Because our silica coupling fibre is incapable of efficiently coupling out mid-IR light, such long wavelengths cannot be directly observed in our experiments; rather, the positions of the mid-IR sidebands shown in Fig.~\ref{fig3}(b) were inferred from the corresponding short-wavelength signals. We expect this issue can be overcome by selectively out-coupling the long-wavelength sidebands using a prism that is transparent in the mid-IR.

In Fig.~\ref{fig3}(c)--(e), we respectively show typical spectra measured from three of our largest devices [cf. Fig.~\ref{fig3}(a)] in more detail. In stark contrast with observations made in silica microsphere systems~\cite{sayson_widely_2017}, each spectrum displays only a single pair of FWM sidebands, with no signatures of parasitic components associated with stimulated Raman scattering or other nonlinear processes. Combined with our other experiments, these results unequivocally show that MgF$_2$ microdisk resonators -- driven with standard low-power telecommunications lasers --  can deliver clean and low-noise parametric sidebands that can be tuned over an entire optical octave.

\begin{figure}[!t]	
\centering
		\includegraphics[width=\linewidth]{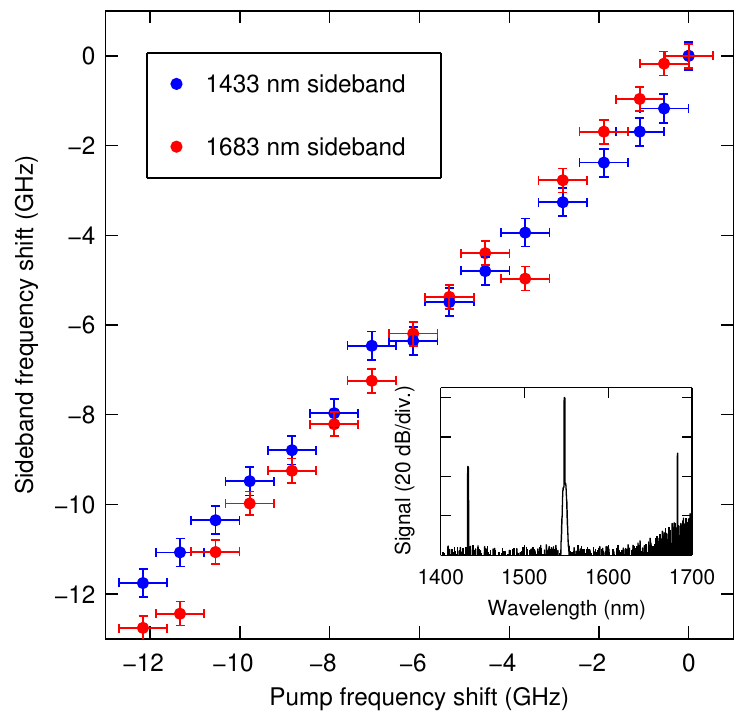}
		\caption{\textbf{Proof-of-concept demonstration of continuous sideband tunability.} Relative frequency shifts of parametric sidebands at approximately $1433~\mathrm{nm}$ (red circles) and $1686~\mathrm{nm}$ (blue circles) as the frequency of the pump (centred at $1550~\mathrm{nm}$) is scanned over 12~GHz. The results were obtained in a resonator whose major radius $R = 265~\mathrm{\mu m}$ and with a pump power of $80~\mathrm{mW}$. The sideband wavelengths were measured using a high-precision wavemeter [see Methods]. Inset shows the entire optical spectrum.}
		\label{fig25}
\end{figure}

We must note that the wideband tunability demonstrated in Figs.~\ref{fig2} and \ref{fig3} is inherently discrete due to the requirement that the pump frequency must always be (almost) resonant with a cavity mode. To achieve continuous tunability, the cavity mode spectrum must be tuned in tandem with the pump frequency, which can be achieved, e.g., by controlling the resonator temperature~\cite{delhaye_octave_2011, xue_thermal_2016} or by applying a mechanical strain on the resonator~\cite{papp_mechanical_2013, schunk_interfacing_2015}. To realise a proof-of-concept demonstration of continuous sideband tunability, we exploit the microresonator's intrinsic thermo-optic nonlinearity~\cite{carmon_dynamical_2004, delhaye_octave_2011}. Here the frequency of an initially blue-detuned pump laser is scanned into resonance, which causes the cavity modes to shift as the build-up of intracavity intensity changes the resonator temperature. By following the thermally-shifted cavity mode, the pump can remain (almost) resonant over an extended range of about 12~GHz, corresponding to about 5,000 cold cavity linewidths. Figure~\ref{fig25} shows the change in frequency of the parametric sidebands as the pump frequency is changed, measured with a high-precision wavemeter [see Methods]. As can be seen, the change in each sideband's frequency follows exactly the change in frequency of the pump, thus demonstrating approximately 12~GHz of continuous tunability. We envisage that the range of continuous tunability can be significantly extended by externally heating the resonator~\cite{xue_thermal_2016}.

\subsection{Discussion}

Our work shows that parametric oscillation in crystalline Kerr microresonators represents a viable route for the generation of widely-tunable laser light. The typical conversion efficiencies observed in our work are still rather small, however, ranging from $10^{-5}$ to $10^{-3}$ for sidebands within the range of our OSAs.  We believe that such a low level of conversion is not a fundamental limit, but arises rather due to un-optimised coupling and attenuation in our detection scheme. Extensive numerical and analytic investigations indicate that, with further optimisation, conversion efficiencies in excess of 10\% to each sideband should be feasible [see Supplementary Information]. These investigations reveal in particular that the out-coupling efficiency plays a very important role in determining the strength of the parametric sidebands outside of the cavity (and hence the conversion efficiency). As noted above, the silica fibre taper used in our experiments is designed to efficiently couple light around 1550~nm into the microresonator, but it does not allow for efficient out-coupling of the large-frequency shift sidebands. We suspect this fact represents the main explanation for the comparatively low conversion efficiencies observed. We expect that selective out-coupling with a mid-IR transparent prism will enable significant improvements to conversion efficiency.

To conclude, we have shown that crystalline Kerr microresonators provide an ideal platform for widely-tunable parametric oscillation. Thanks to their ultra-high finesse, narrow Raman gain spectrum, and attractive dispersion characteristics, clean and low-noise sidebands separated by more than an octave can be generated -- and tuned over hundreds of nanometres -- using a low-power telecommunications pump laser. The ability to fabricate resonators from materials (such as MgF$_2$) that remain transparent deep into the mid-IR presents alluring opportunities for low-cost sources of light in that spectral region. Moreover, the use of materials such as silicon or silicon nitride could allow for the scheme to be realised in an integrated, on-chip format~\cite{moss_new_2013, brasch_photonic_2016, joshi_thermally_2016,griffith_silicon-chip_2015}. We note in this context that tunable mid-infrared generation has recently been achieved by launching intense femtosecond pulses into single-pass silicon nitride \emph{waveguides}~\cite{kowligy_tunable_2018}, alluding to the possibility of fabricating on-chip resonators with suitable dispersion characteristics.

\section*{Methods}

\footnotesize

\subparagraph*{\hskip-10pt Phase-matching considerations.}
The nonlinear dynamics of a driven Kerr (micro)resonator can be modelled using the generalised Lugiato-Lefever equation (LLE)~\cite{coen_modeling_2013}. A standard linear stability analysis can be used to determine the per-roundtrip amplitude gain $g(\Omega)$ experienced by a small sideband perturbation with frequency shift $\Omega$ superimposed on top of a cw steady-state solution of the system~\cite{sayson_widely_2017, haelterman_additive-modulation-instability_1992}:
\begin{equation}
\label{gb}
g(\Omega) = -\alpha + \sqrt{\gamma^2L^2P_0^2 - (\delta_0-D_e(\Omega,\omega_\mathrm{p})L - 2\gamma L P)^2}.
\end{equation}
Here, $\alpha$ is half of the power lost per round-trip and $D_e(\Omega,\omega_\mathrm{p})$ describes the dispersion-induced phase mismatch of the parametric interaction. This latter parameter depends only on the even orders of dispersion:
\begin{equation}
D_e(\Omega,\omega_\mathrm{p}) = \sum_{k\geq1} \frac{\beta_{2k}}{(2k)!}\Omega^{2k},
\end{equation}
where $\beta_k = d\beta/d\omega|_{\omega_\mathrm{p}}$ is the $k^{\mathrm{th}}$ derivative of the the mode propagation constant $\beta(\omega)$ evaluated at the pump frequency $\omega_\mathrm{p}$.

Parametric gain can be achieved at frequencies for which \mbox{$g(\Omega)>0$}. It is easy to show that the gain is maximised at the phase-matched frequency that satisfies
\begin{equation}
\label{PM2}
D_e(\Omega,\omega_\mathrm{p})L + 2\gamma L P - \delta_0 = 0.
\end{equation}
Truncating the dispersion operator $D_e(\Omega,\omega_\mathrm{p})$ to its first two terms (fourth order in dispersion) gives the simplified phase-matching Eq.~\eqref{PM} introduced in the main text. While this truncation may not be entirely accurate for some of the larger frequency shifts observed in our experiments, we expect the resulting deviations to be qualitative only. Indeed, we note that the phase-matching curves observed experimentally are well-fitted by the simplified phase-matching Eq.~\eqref{PM}. These curves are also in reasonable qualitative agreement with resonator dispersion characteristics extracted from finite-element simulations. In this context, we note that quantitative agreement between our experiments and finite-element simulations is not to be expected due to (i) uncertainties in the bulk refractive index of MgF$_2$ over the extreme wavelength ranges accessed in our experiments and (ii) uncertainties in the precise morphology of our resonators. Regardless, our finite-element simulations readily predict the general trends that underpin our work, namely the shifting of the resonator ZDW to longer wavelengths with decreasing major radius $R$.

Steady-state parametric oscillations can be expected close to the phase-matched frequency that satisfies Eq.~\eqref{PM2}. An additional restriction stems from the requirement that the parametric sidebands must be resonant with the cavity. Thanks to the non-zero gain bandwidth described by Eq.~\eqref{gb}, it is generally possible to satisfy both conditions simultaneously. However, as noted in ref.~\cite{sayson_widely_2017}, the parametric gain bandwidth tends to decrease with increasing sideband frequency shift. As a consequence, for sufficiently large frequency shifts, the parametric gain may fall in between cavity modes, prohibiting oscillation or reducing its efficiency.

The sideband wavelengths can be tuned by controlling the pump frequency. This changes the dispersion operator $D_e(\Omega,\omega_\mathrm{p})$ and hence the frequency shift $\Omega$ that satisfies the phase-matching condition. For example, when truncating the dispersion at fourth-order, a small pump frequency shift $\delta\omega$ transforms the group-velocity dispersion coefficient as
\begin{equation}
\beta_2 \rightarrow \beta_2 + \beta_3\delta\omega + \frac{\beta_4}{2}\delta\omega^2.
\end{equation}

\subparagraph*{\hskip-10pt Resonator modelling.}
To model the dispersion characteristics of our resonators, we use a commerical finite-element software (COMSOL Multiphysics) to calculate the resonant eigen-frequencies $\omega_m$ of the fundamental TE mode of the resonator. From the resonant frequencies, we obtain the mode propagation constant $\beta(\omega)$ by using the resonance condition $\beta(\omega_m)L = 2\pi m$. A Taylor series expansion at a pump frequency $\omega_\mathrm{p}$ of interest finally yields the dispersion coefficients $\beta_k$.

We note that an alternative description of dispersion -- often encountered in studies of microresonators -- involves expanding $\omega_m$ (rather than $\beta(\omega)$) as a Taylor-series around a particular resonance frequency $\omega_{m_0}$:
\begin{equation}
\omega_m = \omega_{m_0} + \sum_{n\geq1}\frac{D_n}{n!}(m-m_0)^n,
\end{equation}
where $D_n$ stand for the Taylor series coefficients. While the use of this description may seem more natural, it is worth noting that the parametric sidebands are symmetrically detuned with respect to the pump frequency $\omega_\mathrm{p}$ which does not in general coincide with a resonant frequency (e.g. $\omega_{m_0}$). As a consequence, while the phase-matching conditions can be expressed in terms of the Taylor series coefficients $D_n$, the odd-orders of dispersion no longer cancel out in general, resulting in slightly more complex expressions.

\subparagraph*{\hskip-10pt Additional experimental details.}
The driving laser used in most of our experiments is a commercially available C-band laser (New Focus, Velocity TLB-6728) with a narrow linewidth of about 70~kHz. The laser is amplified by a C-band EDFA, and then filtered by a tunable 1~nm passband filter to remove the majority of the amplifier's amplified spontaneous emission noise. Another laser (New Focus, Velocity TLB-6730), tunable in the L-band ($1550~\mathrm{nm} - 1630~\mathrm{nm}$), was used in combination with an L-band EDFA to extend the pump wavelength range up to 1590~nm for the measurements shown in Fig.~\ref{fig3}(b).

To measure the linewidth of the resonator modes, we phase modulate the driving laser to provide an absolute frequency reference, and then scan the laser frequency across the mode of interest in the low-power linear regime. The finesse is obtained by dividing the free-spectral range estimated from the resonator geometry with the measured resonance linewidth.

Our resonators support several different mode families, only some of which exhibit dispersion characteristics suitable for large-frequency shift sidebands. To identify appropriate modes, a part of the light coupled out from the resonator is passed through an optical filter that transmits only wavelengths below 1520~nm. Monitoring the signal transmitted by this filter as the laser wavelength is scanned allows us to straightforwardly single out modes generating large-frequency shift sidebands. Once suitable modes have been identified, we continuously tune the laser into resonance from the blue-detuned side so as to leverage thermal locking~\cite{carmon_dynamical_2004}.

The linewidth measurements reported in Fig.~\ref{fig2}(e) were obtained using a delayed self-heterodyne measurement. Here the sideband of interest is first spectrally isolated using a tunable optical filter, then divided into two parts. One part is frequency-shifted by 80~MHz with an acousto-optic modulator, while the other part passes through 30~km of fibre so as to de-correlate the two beams. Superimposing the beams on a photodetector allows the linewidth to be estimated from the recorded beat signal.

Two optical spectrum analysers and an FTIR spectrometer were used to gather the data shown in Fig.~\ref{fig3}. The first OSA (Yokogawa AQ6370D) measures from 600~nm to 1750~nm, the second OSA (Yokogawa AQ6375B) measures from 1200~nm to 2400~nm, and the FTIR spectrometer (Bristol Instruments 771B; note that this device can operate as a high-precision wave meter or an FTIR spectrometer) was used to record spectra of sidebands with wavelengths above 2400~nm. Because the dynamic range of the FTIR spectrometer is limited to 45~dB, a bandpass filter with a centre wavelength of about 2500~nm and a pass-band of 500~nm was placed just before the spectrometer so as to filter out the strong pump line. The two different OSAs and the FTIR were calibrated with a signal at 1550~nm so as to construct the spectra shown in Fig.~\ref{fig3}(c)--(e).

To demonstrate continuous sideband tunability [cf. Fig.~\ref{fig25}], we first tune the pump laser into an appropriate resonance so as to generate parametric sidebands. We then isolate the pump and the sidebands using appropriate spectral filters, and measure their wavelengths using a high-precision wave meter (Bristol Instruments 771B) with wavelength accuracy of 0.75 parts-per-million. We then continuously decrease the frequency of the pump and re-measure all the wavelengths. This procedure is repeated for as long as the thermo-optic nonlinearity allows the pump to remain thermally locked to the resonance. We must emphasise that, because our wave meter only allows one wavelength to be measured at a time, there is a short (approximately 15 second) delay between the measurement of the pump and the sideband wavelengths. During this delay, the pump wavelength can exhibit small fluctuations, giving rise to an additional uncertainty in the precise pump wavelength for which each sideband wavelength was measured. This uncertainty is included in the horizontal error bars in Fig.~\ref{fig25}.

\section*{Acknowledgments}

\noindent This work was supported by the Marsden Fund, Rutherford Discovery Fellowships, and James Cook Fellowships of the Royal Society of New Zealand.

\section*{Author Contributions}

\noindent N.S. performed all the experiments. T.B. and V.N. performed numerical modelling of resonators, and helped perform the delayed self-heterodyne linewidth measurements. L.T. and H.S. fabricated the resonators. M.E. and S.C. contributed to theoretical interpretation of the results. M.E. and S.G.M. wrote the manuscript. All authors contributed to discussing and interpreting the results.

\section*{Data availability}
\noindent The data that support the plots within this paper and other findings of this study are available from the corresponding author upon reasonable request.

\section*{Competing financial interests}

\noindent The authors declare no competing financial interests.


\end{document}